\documentclass[manuscript]{acmart}

\AtBeginDocument{%
  }

\acmSubmissionID{7336}

\usepackage{enumitem}
\usepackage{graphicx}
\usepackage{colortbl}
\usepackage{array}
\usepackage[table,xcdraw]{xcolor}
\usepackage{multicol}
\usepackage{multirow}
\usepackage{booktabs}
\usepackage{arydshln}
\usepackage{ragged2e}
\usepackage{subcaption}
\usepackage{url}
\usepackage{amsmath}
\usepackage{mathtools}
\usepackage{siunitx}
\usepackage{longtable}
\usepackage[most]{tcolorbox}
\usepackage{todonotes}
\usepackage{balance}
\usepackage[most]{tcolorbox}

\usepackage{booktabs}
\usepackage{graphicx} 
\usepackage{array}

  \usepackage{booktabs}
  \usepackage{array}
  \usepackage{multirow}  
  \usepackage{pifont}
  \newcommand{\yes}{\ding{52}}   
  \newcommand{\no}{\ding{56}}    

\usepackage{xurl}            

\newtcolorbox{rqsummarybox}[2][]{
  colback=gray!5,
  colbacktitle=gray!20,
  colframe=black!50,
  coltitle=black, 
  boxrule=0.6pt,
  title={Summary for #2},
  fonttitle=\bfseries\itshape,
  boxsep=1ex,
  left=1ex,
  right=1ex,
  top=0.8ex,
  bottom=0.8ex,
  breakable,
  #1
}

\makeatletter
\renewenvironment{quote}
  {\list{}{\leftmargin=2.1\parindent \rightmargin=2.1\parindent}%
   \item\relax}
  {\endlist}
\makeatother

\newcommand{\paragraphb}[1]{\vspace{0.07in}\noindent{\bf #1.} }

\begin{document}

\title{How Does Perceived Empathy From AI Chatbots Shape Users’ Privacy Attitudes, Behaviors, and Needs?}

\author{Jiaxun Cao}
\email{jessie.cao@duke.edu}
\affiliation{%
  \institution{Duke University}
  \country{USA}
}

\author{Hiba Laabadli}
\email{hiba.laabadli@duke.edu}
\affiliation{%
  \institution{Duke University}
  \country{USA}
}

\author{Yuntong Xue}
\email{yx3043@nyu.edu}
\affiliation{%
  \institution{New York University}
  \country{USA}
}

\author{Jana Schaich Borg}
\email{js524@duke.edu}
\affiliation{%
  \institution{Duke University}
  \country{USA}
}

\author{Pardis Emami-Naeini}
\email{pardis@cs.duke.edu}
\affiliation{%
  \institution{Duke University}
  \country{USA}
}


\begin{abstract}
\label{sec:abstract}
Prior research suggests that some dimensions of AI empathy may increase users' disclosure to chatbots, posing growing privacy risks as chatbots become more capable of empathic interaction. Yet this work often conceptualizes AI empathy narrowly, leaving users’ own definitions and broader privacy experiences underexplored. We interviewed 21 AI chatbot users about what AI empathy meant to them and how it shaped their privacy attitudes, behaviors, and needs. Participants characterized empathy through five components---understanding, resonating, caring, expressing, and helping. Most participants viewed resonating and caring as essential components of human empathy but absent from AI empathy. Yet AI's contextual, in-depth understanding and helpfulness still encouraged highly sensitive disclosure. This highlights that AI empathy can endanger users' privacy without necessarily making AI seem human. Participants also desired varied forms and levels of empathy, each associated with distinct privacy needs. We offer actionable recommendations for designing privacy-enhancing mechanisms for empathic AI interactions.


\end{abstract}

\maketitle

\section{Introduction} \label{sec:intro}

Generative AI (GenAI) chatbots\footnote{For the remainder of this paper, we use AI chatbots to refer to GenAI chatbots, or conversational agents powered by large language models.} have rapidly expanded their capabilities, including their ability to produce responses that users perceive as highly empathic. A growing body of research suggests that users may experience AI chatbot responses as empathic---sometimes even more so than human-authored responses~\cite{sorin2024large, ayers2023comparing, liu2023leveraging, do2020clinical, shen2024empathy}. Empathy is widely regarded as beneficial across domains such as healthcare~\cite{schmidmaier2025using, jeong2023robotic, dekeyser2009empathy, elliott2011empathy, elliott2018therapist}, emotional support~\cite{schmidmaier2025using, fitzgerald2010active, burleson2003experience, verhofstadt2016role, jones2019impact}, and customer experience~\cite{chen2024empathy, casas2021enhancing, chin2020empathy, guo2021shing, hu2018touch, janson2023leverage}. In Human-Computer Interaction (HCI), empathy has also long been considered a desirable quality and design goal for improving user engagement~\cite{gencc2024situating, huang2026beyond}. However, the expanding empathic capabilities of AI chatbots have introduced safety and ethical risks\footnote{In this paper, we use the term ``AI empathy'' to refer to interactions with general-purpose GenAI systems that users \textit{perceive} or \textit{experience} as empathic. We do not attempt to assess whether these systems possess empathy of any kind.}. For example, AI sycophancy---a key empathic cue in which a chatbot persistently agrees with users to maximize engagement~\cite{cheng2025elephantmeasuringunderstandingsocial}---can contribute to delusion~\cite{wsj2025delusionalspirals, nytimes2025spiraling, psychtoday2025aipsychosis, longreads2025divorce}, addiction~\cite{kelly2020your}, and emotional dependency~\cite{visioemotional}. Widespread concern has also emerged about users' inability to distinguish AI from humans, particularly among vulnerable users who turn to AI for companionship~\cite{nytimes2025spiraling}.

While the safety and ethical implications of AI empathy have received growing attention, its privacy implications remain underexplored. Two adjacent lines of research offer relevant, though indirect, evidence that AI empathy might pose privacy risks to users with its emotional capacities and humanlike-ness. First, studies have found that users disclose more when interacting with chatbots that exhibit greater emotional expressiveness~\cite{lee2020designing, lee2020hear, loveys2020effect}. Second, more recent research has shown that AI chatbots’ anthropomorphic cues and perceived human-likeness can encourage more sensitive self-disclosure~\cite{zhang2024s, zhan2025malicious}.

However, emotional capacity and human-likeness do not fully capture empathy, which is a complex, multidimensional construct. Prior research distinguishes cognitive empathy (i.e., understanding another person's thoughts or emotions) from affective empathy (i.e., sharing those emotions), and has proposed further dimensions~\cite{gencc2024situating, gallese2003roots}. Lay conceptions of empathy may be even broader and more nuanced than these academic constructs~\cite{hall2021laypeople}. Yet existing work has neither identified which dimensions users associate with AI empathy nor systematically examined how these dimensions shape their privacy experiences. Distinguishing these dimensions is important because they may shape users’ privacy perceptions and behaviors through different mechanisms, potentially creating distinct privacy risks that require targeted protections. The privacy implications of AI empathy may also extend beyond the level of self-disclosure to users’ privacy attitudes, risk perceptions, protective behaviors, and needs for privacy support. We therefore aim to move beyond examining AI empathy solely in terms of emotional capacity or human-likeness. Specifically, we systematically investigate how users conceptualize AI empathy in their own words, how its distinct dimensions shape their privacy experiences, and what privacy protections they need across different forms and levels of empathic interaction. Toward this end, our work asks three research questions (RQs):

\begin{itemize}
    \item \textbf{RQ1}: How do users conceptualize and perceive empathy in interactions with AI chatbots, and how, if at all, do they distinguish AI empathy from human empathy?
    \item \textbf{RQ2}: What are users' privacy perceptions and behaviors when perceiving different dimensions of empathy from AI chatbots?
    \item \textbf{RQ3}: What are users' expectations toward privacy-enhancing empathic AI interactions?
\end{itemize}

To answer these questions, we conducted semi-structured interviews with 21 U.S.-based AI chatbot users to examine their conceptualizations of AI empathy and their privacy perceptions, behaviors, and needs in interactions they perceived as empathic. Participants primarily characterized empathy through five components---understanding, resonating, caring, expressing, and helping. Although participants varied in the components they emphasized or omitted, most distinguished AI empathy from human empathy and did not believe that AI could emotionally resonate with or care about them as humans do (\textbf{RQ1}). Yet participants reported that AI chatbots’ contextual, in-depth understanding and helpfulness still led them to disclose more sensitive information than they otherwise would have shared. These findings highlight that AI empathy can pose privacy risks without necessarily making AI seem human (\textbf{RQ2}). In response to these risks, participants called for greater control over when and how AI chatbots express empathy, alongside safer data practices for empathic AI interactions (\textbf{RQ3}). Grounded in these findings, we discuss design opportunities, recommend data practices, and identify directions for future research to support context-aware, privacy-enhancing empathic AI interactions.

Our work makes three contributions to the HCI community: 1) an empirically grounded, user-centered characterization of AI empathy across five components; 2) empirical insights into how these components shape users’ privacy experiences; and 3) actionable recommendations for providing contextually appropriate empathy controls and implementing privacy-enhancing data practices in empathic AI interactions.

\section{Related Work} \label{sec:background}
\paragraphb{Cross-disciplinary conceptions of empathy} Studying the impact of empathy on users' attitudes and behaviors requires an understanding of how empathy is defined. Multiple disciplines, including psychology, neuroscience, sociology, philosophy, and others, debate about the definition of empathy~\cite{eklund2021toward, blomster2025empathy, batson2009these, blystad2022empathy,cuff2016empathy, sulzer2016assessing, hall2022empathy, cuadra2024illusion, anccel2006developing, paiva2017empathy, gencc2024situating, borg2024required}. These debates are not fully resolved, but there is nonetheless an emerging consensus that empathy entails some combination of understanding, feeling, and sharing another's world while maintaining self-other differentiation~\cite{eklund2021toward}. Empathy is often discussed in research as a multi-dimensional umbrella covering three main components: cognitive, affective, and motivational empathy~\cite{gencc2024situating, gallese2003roots}. Cognitive empathy refers to the recognition and understanding of another person's thoughts or emotional states~\cite{cuff2016empathy, li2025exploring, chen2024empathy, drouet2022empathy, kouprie2009framework, feshbach1975empathy}. Affective empathy refers to the sharing or experiencing of another's emotions, while generally maintaining self-other differentiation ~\cite{sorin2024large, blair2008fine}. Motivational empathy~\cite{vaish2024sharp, gallagher2024pattern} involves feelings of concern for another, accompanied by a genuine motivation to take actions~\cite{borg2024required, perry2023ai}. At the same time, lay people have been shown to use the term \textit{empathy} in a much more diverse way than academic constructs suggest~\cite{hall2021laypeople, blomster2025empathy}. Hall et al.~\cite{hall2021laypeople}, replicated by a 2025 study~\cite{blomster2025empathy}, found that participants varied widely in how they endorsed items drawn from scales based on academic definitions of empathy. This suggests it is important to first understand how a given population defines empathy when interpreting their responses to perceived empathy. Accordingly, we began by asking participants how they define empathy in their day-to-day lives, and then how they define empathy specifically in the context of interacting with AI.

\paragraphb{Societal-wide benefits of empathy} The growing use of empathic capabilities in AI chatbots is grounded in extensive evidence that empathy is beneficial across many critical domains, including physiological and psychological health care, emotional and social support, user and customer experience, and the promotion of prosocial behavior. In mental health care, empathy is associated with stronger therapeutic alliance and better treatment outcomes~\cite{schmidmaier2025using, jeong2023robotic, dekeyser2009empathy, elliott2011empathy, elliott2018therapist, feller2003importance, herbener2024narrative, nienhuis2018therapeutic, wampold2015important, wampold2015great, watson2016role}. In contexts where people seek emotional and social support, empathy provides active listening~\cite{schmidmaier2025using, fitzgerald2010active, burleson2003experience, verhofstadt2016role, jones2019impact}, perspective taking~\cite{cuff2016empathy, decety2005perspective}, and non-judgmental behavior~\cite{bae2021social, herbener2025lonely, narain2020promoting}. Consequently, empathy is a desirable quality in customer service and user experience design, as evidenced by higher satisfaction scores when users perceive interactions as empathic~\cite{chen2024empathy, casas2021enhancing, chin2020empathy, guo2021shing, hu2018touch, janson2023leverage}. In addition, extensive research identifies empathy as a psychological mechanism that promotes prosocial behavior at scale~\cite{li2025exploring, eisenberg2002empathy, eisenberg1987relation, oliveira2021towards, hoffman2008empathy, yin2023empathy, chen2024empathy, crowley2011behavioral, sierksma2015group, taylor2019accountability, van2015emotional, roberts1996empathy}. Taken together, existing research shows that empathy can produce societal-level benefits, making it a valuable quality for user-facing AI systems to demonstrate.

\paragraphb{Users' perceptions of empathic capabilities in AI chatbots} Users often perceive empathy from AI chatbots through human-like communication that leverages three types of cues: (1) identity, (2) verbal, and (3) nonverbal cues. \emph{Identity} cues include information such as the chatbot's avatar, name, and gender~\cite{berry2005evaluating, cowell2005manipulation, janson2023leverage}. \emph{Verbal} cues refer to the use of words and phrases~\cite{bickmore2005establishing, janson2023leverage, sah2015effects, seeger2018designing}, e.g.,  acknowledging users' feelings. \emph{Nonverbal} cues refer to non-linguistic behaviors in communication, such as chatbots' use of emoticons~\cite{derks2008emoticons}. These three types of cues have been documented in prior work to influence users' perceptions of chatbots’ empathic capabilities, thereby shaping interaction dynamics~\cite{schmidmaier2025using, li2025exploring, seeger2018designing}. For example, many users have noted ChatGPT's frequent use of emojis, leading to increased perceptions of the chatbot's human-like nature~\cite{nindzsaaa2025emoji}.

A growing body of work suggests that chatbot responses are often perceived as more empathic than human-generated responses~\cite{sorin2024large, ayers2023comparing, liu2023leveraging, do2020clinical, shen2024empathy}. For example, Ayers et al.~\cite{ayers2023comparing} compared health care professionals' ratings of ChatGPT-generated responses to patients' questions with those of human physicians and found that ChatGPT responses were preferred in 78.6\% of cases and had responses participants rated as empathetic or very empathetic almost 10 times more than those provided by humans.

At the same time, sometimes AI chatbots also fail to provide desirable empathy. For example, Chen et al.~\cite{chen2023llm} used ChatGPT to simulate role-play scenarios between psychiatrists and patients and noted that ChatGPT sometimes forgot initial instructions and often produced repetitive empathic phrases. Similarly, other researchers have also observed repetitive patterns in AI chatbots' empathic expressions~\cite{yang2022effect, zhao2023chatgpt}. Schaich Borg and Read~\cite{borg2024required} suggest that users might prefer chatbots' empathic responses in some contexts but find those same responses ``annoying, eerie, untrustworthy, or unethical,'' in other contexts, calling for more research into users' preferences for empathic qualities in AI chatbots. 

\paragraphb{Ethical and safety risks of AI empathy} The surge in empathic AI chatbots, coupled with their expanding empathic capabilities, raises pressing risks, notably the potential for deception and manipulation~\cite{shteynberg2024does, williams2024targeted, visioemotional, nytimes2025spiraling, cuadra2024illusion, adam2025supportive, washpost2025messingmind}. In illustration, evidence suggests that one tactic for showing empathy and thus maximizing engagement is to continually agree with users, a phenomenon known as AI sycophancy~\cite{cheng2025elephantmeasuringunderstandingsocial}. This tendency has resulted in negative consequences. Specifically, OpenAI acknowledged in a blog post in April 2025 that an updated version of GPT-4o had begun ``validating doubts, fueling anger, urging impulsive actions, or reinforcing negative emotions in ways that were not intended''~\cite{openai2025sycophancy}. AI sycophancy can also contribute to user delusion~\cite{wsj2025delusionalspirals, nytimes2025spiraling, psychtoday2025aipsychosis, longreads2025divorce}. McCoy et al. tested 38 major AI models and found that GPT-4o affirmed claims indicative of possible psychosis 68\% of the time~\cite{nytimes2025spiraling}. Even users without psychosis or underlying mental health conditions can get addicted to such constant validation, as researchers suggest that social validation---a key strategy for signaling empathy---is linked to addiction and emotional dependency~\cite{kelly2020your}. Consequently, some experts worry about users seeking emotional support from AI chatbots, noting that such blind validation without constructive input can narrow users' worldviews and create a new type of echo chamber~\cite{moore2025expressing, longreads2025divorce, psychtoday2025therapistrisks, guardian2025notsafe}.

Moreover, another ethical question arises regarding whether and under what conditions users can distinguish AI-simulated empathy from genuine human empathy~\cite{shteynberg2024does, cuadra2024illusion}. As argued by~\cite{shteynberg2024does}, if AI does not experience empathy internally, the deployment of such inauthentic empathy---systems that appear to care but do not, in reality---may be considered deception. Researchers have also called for further work on users’ preferences regarding whether AI chatbots must internally share or experience emotions to deliver desirable empathy~\cite{borg2024required, inzlicht2024praise}.

\paragraphb{Lack of privacy research regarding empathic AI} While ethical and safety concerns dominate media and scholarly discourse on empathic AI interactions, the privacy implications remain underexplored. Prior work has primarily focused on investigating the level of self-disclosure when users interact with chatbots equipped with some human-like features~\cite{schmidmaier2025using, laban2023opening, lee2020hear, loveys2020effect, bae2021social, martelaro2016tell, lee2020designing, ho2018psychological, lucas2017reporting, lee2024influence}. Specifically, research has observed higher levels of self-disclosure when users interact with more emotionally expressive chatbots~\cite{laban2023opening, lee2020hear, loveys2020effect, bae2021social, martelaro2016tell, lee2020designing, lucas2017reporting, lee2024influence}. For instance, Lee et al.~\cite{lee2020designing, lee2020hear} found that participants disclosed more when speaking with a chatbot that talked about its own feelings.

Still, \textbf{there is a critical lack of research on how perceived empathy in AI chatbots shapes users' privacy risk perceptions, decision-making, behaviors, and mitigation practices.} These understandings are crucial for informing the design of user-centric privacy solutions. This research agenda is particularly timely and important in an era of rapidly advancing AI capabilities, as users increasingly turn to empathic AI for highly sensitive purposes~\cite{li2024human}. Privacy, therefore, is a fundamental issue that warrants closer examination. Our work thus contributes by providing an empirical account of users' preferences and offering actionable suggestions for the design of empathic AI interactions that are ethical, safe, and privacy-preserving.
\section{Methodology} \label{sec:methods}
To elicit empirical insights into how perceived empathy from AI chatbots shapes users' privacy attitudes, behaviors, and needs, we conducted semi-structured interviews with 21 users of AI chatbots.

\paragraphb{Recruitment} 
From June 2025 to August 2025, we recruited participants on the Prolific crowdsourcing platform\footnote{\href{https://www.prolific.com/}{https://www.prolific.com/}}. To mitigate social desirability~\cite{grimm2010social} and demand characteristics~\cite{mccambridge2012effects}, we did not include any words related to privacy or security in our recruitment and screening materials (see Appendix~\ref{sec:appendix:artifacts}). We advertised our study as an investigation of users' experiences and perspectives toward empathy in AI chatbots. Prolific users interested in our study were redirected to the Qualtrics survey platform\footnote{\href{https://www.qualtrics.com/}{https://www.qualtrics.com/}}, where we hosted our pre-screening survey. Our study obtained approval from our institution's review board (IRB). All participants in the pilot and the main study provided their informed consent before participating. Information about the recruited participants and their AI chatbot usage is detailed in Sec~\ref{sec:results}.

\paragraphb{Eligibility and pre-screening} To participate in our interview study, participants must (1) be at least 18 years old, (2) live in the United States, (3) have used an AI chatbot at least once, (4) have \textit{perceived} empathy from an AI chatbot, and (5) have a task approval rate of at least 95\% on the Prolific platform. To find eligible participants and to recruit a diverse sample of participants, in the pre-screening survey, we collected information about their AI chatbot usage (all the chatbots used, purposes for usage, and usage frequency) and general demographics (e.g., age, gender, and education). Because one of our goals was to elicit rich empirical insights into how perceptions of AI empathy shaped users' privacy experiences, we required participants to have used AI chatbots at least once and to have perceived empathy from an AI chatbot at least once, as specified in the screening survey and interview questions. In addition, we sought to capture perceptions of empathy beyond personal and emotional contexts and therefore purposively recruited participants whose AI chatbot use spanned a diverse range of use cases (see Table~\ref{tab:participants}). Requiring a minimum level of AI usage and perceived empathy in AI chatbot interactions helps mitigate potential hypothetical bias~\cite{harrison2008experimental}, wherein participants report attitudes and behaviors that diverge from their actual practices. All the participants, regardless of their eligibility, were compensated at an hourly rate of 21.32 USD through the Prolific platform. The median time for completion was about 2 minutes. The recruitment and pre-screening materials are included in the Appendix~\ref{sec:appendix:artifacts}.

\paragraphb{Pilot interviews} Prior to the formal interview study, we conducted pilot interviews with three participants recruited through the first author's personal connections. The three participants all qualified for our eligibility criteria. The purpose of the pilots was to assess and improve the clarity of the interviews and ensure that the interviews could be completed within 45 minutes to 1 hour. After each pilot, the researchers revised the interview protocol to address any confusing questions. We applied minor wording edits to the interview questions to improve clarity and flow of the script, and removed redundancies to shorten the interview procedure. No improvements or changes were made after the two pilot interviews.

\paragraphb{Interview procedure} We reached out to each participant via Prolific messaging one day before the interview to confirm their attendance. We conducted 21 interviews via the Zoom remote communication platform~\footnote{\href{https://zoom.us/}{https://zoom.us/}}. We reached data saturation~\cite{saunders2018saturation} at the seventeenth interview, i.e., no additional insights emerged. Aligned with qualitative best practices~\cite{francis2010adequate}, we then conducted four more interviews without identifying any new insights.

To address our three research questions (Sec~\ref{sec:intro}), our interview protocol (see Appendix~\ref{sec:appendix:interview}) consists of five main sections:

\begin{itemize}[leftmargin=0pt]
    \item \textbf{Section 1: AI chatbot usage.} After introducing the purpose of the study and the research team, we asked participants about their usage of AI chatbots, including the chatbots they have used, the purposes of usage, usage frequency, and any chatbots they no longer use and their reasons behind these decisions.
    \item \textbf{Section 2: Definitions, understandings, and experiences of empathy from AI chatbots (RQ1).} Next, we asked participants about their definitions of empathy first in a general sense and then AI empathy in particular. Then, we asked about participants' experiences perceiving empathy from AI chatbots. We then asked participants about the perceived benefits and harms of AI empathy, as well as the contexts in which participants want or do not want empathy from an AI chatbot.
    \item \textbf{Section 3: Privacy attitudes and concerns toward perceived empathy from AI chatbots (RQ2).} After surfacing participants' general perceptions of empathy from AI chatbots, we shifted our focus to security and privacy specifically. We asked the participants about their security and privacy concerns regarding perceived empathy from AI chatbots, if any. We also asked the participants if they shared information of different levels of sensitivity with AI chatbots when they perceived empathy from AI chatbots. We then asked about participants' mitigation strategies.
    \item \textbf{Section 4: Expectations regarding privacy-enhancing empathic AI interactions (RQ3).} We asked participants about their expectations around what features and capabilities they want to better protect their privacy when engaging in empathic AI interactions.
\end{itemize}

\paragraphb{Qualitative data analysis} All the interviews were audio-recorded and transcribed using the Zoom transcription service for data analysis. The first author proofread all the transcripts before the analysis, ensuring accuracy and readability. Three researchers conducted a thematic analysis using the inductive approach~\cite{saldana2021coding}. Our qualitative analysis consists of three iterative stages:

\begin{itemize}[leftmargin=0pt]
    \item \textbf{Stage 1: Independent coding.} Three researchers independently coded three identical transcripts and generated their own codebook.
    \item \textbf{Stage 2: Establishing an initial codebook.} Three researchers presented their codebooks to each other. After discussing, merging, and reconciling the codes, three researchers agreed on an initial codebook.
    \item \textbf{Stage 3: Iterating the codebook.} Three researchers split the remaining transcripts and coded them using the merged codebook based on consensus. Two researchers coded each transcript. During this stage, new codes were added, removed, and merged. Through several meetings with the entire research team, the researchers discussed their codes and resolved all disagreements. Since all the transcripts were double-coded and disagreements were resolved through the meetings, calculating inter-coder reliability was not necessary~\cite{mcdonald2019reliability}. The final codebook is included in Appendix~\ref{sec:appendix:artifacts}.
\end{itemize}

\paragraphb{Limitations} As is common in qualitative studies, our sample size was relatively small and therefore generalizability might be limited in several ways. First, all of our participants were recruited via Prolific, which introduced selection bias~\cite{scribbr_selection_bias} toward technically savvy users. To mitigate this bias, we had two pre-screening questions, one about participants' education level and the other about their technical background. When selecting participants, we aimed for a balanced distribution among participants' educational backgrounds and technical expertise. Additionally, we avoided including privacy and/or security-related words in our recruitment and pre-screening materials to mitigate attracting participants who are more concerned about these topics. Second, the study relied on participants' self-reported information, which may be subject to social desirability bias~\cite{grimm2010social}. Third, we recruited participants who had perceived empathy from AI chatbots in order to elicit richer accounts, but this choice may have excluded the perspectives of users who do not perceive AI chatbots as empathic. We sought to mitigate this bias by purposively recruiting participants with a wide range of use cases, from personal to work contexts where empathy levels may vary. Nevertheless, our findings may not be generalizable to all users of AI chatbots.

In addition, because little prior research examines how lay users conceptualize empathy from AI chatbots, we first sought to capture participants' lived experiences qualitatively, along with their related privacy preferences. Future work should complement this account with quantitative studies, such as lab and field experiments, that examine the relationship between perceived empathy and privacy risks. Our exploratory findings, including the taxonomy of AI empathy perceptions, offer hypotheses for such work to test.

\paragraphb{Ethics statement}
Our study received approval from our Institutional Review Board (IRB). All participants in the pilot and main interviews provided informed consent before participation. During interviews, we reminded participants that they could skip any question, pause, or stop at any time without penalty, and that they could keep their camera off. Participants were compensated at a flat rate of 20 USD, aligned with institutional and platform norms. Prior to analysis, the first author proofread and de-identified all raw data by removing names, Prolific IDs, and other identifying information. We used randomly assigned participant codes (P1-P21) in analysis and reporting. Only the research team had access to the raw data.
\section{Results} \label{sec:results}
\begin{table*}[t]
\centering
\resizebox{\textwidth}{!}{%
\footnotesize{
\begin{tabular}{l p{4cm} p{8cm} l l l p{2cm}}
\toprule[1.1pt]
\textbf{ID} & \textbf{Chatbot(s) Used} & \textbf{Purposes of Usage} & \textbf{Frequency of Usage} & \textbf{Age} & \textbf{Gender} & \textbf{Ethnicity} \\
\midrule
P01 & ChatGPT, Google Bard, Meta Llama 2, Microsoft Bing AI, Perplexity, Other: Deepseek & Therapy/companionship, Organizing my life, Finding life purpose, Enhanced learning, Generating ideas, Fun and nonsense, Creativity, Healthier living, No particular task, I just wanted to see what it was like & More than once a day & 35 - 44 & Female & Black or African American \\
P02 & ChatGPT & Therapy/companionship, Healthier living, No particular task, I just wanted to see what it was like & Once a week & 45 - 54 & Male & White \\
P03 & ChatGPT, Claude, Google Bard, Google Gemini, Microsoft Bing AI, Other: Deepseek, Minstral & Therapy/companionship, Organizing my life, Finding life purpose, Enhanced learning, Generating code (for pros), Generating ideas, Fun and nonsense, Creativity, Healthier living & Once a day & 45 - 54 & Male & Asian \\
P04 & ChatGPT, Claude, Google Gemini, Grok & Therapy/companionship, Enhanced learning, Generating ideas, Fun and nonsense, Creativity & More than once a day & 35 - 44 & Male & Asian \\
P05 & Character.ai, ChatGPT, Google Gemini & Therapy/companionship, Generating ideas, Fun and nonsense, Creativity & More than once a day & 18 - 24 & Female & Black or African American, White \\
P06 & Character.ai, ChatGPT, Google Bard, Meta Llama 2, Microsoft Bing AI & Therapy/companionship, Enhanced learning, Generating code (for pros), Generating ideas, Creativity, Healthier living & Once a week & 35 - 44 & Male & Asian \\
P07 & ChatGPT, Claude, GitHub Copilot, Google Gemini & Therapy/companionship, Enhanced learning, Generating ideas, Fun and nonsense, Creativity, Healthier living & Once a day & 45 - 54 & Female & White \\
P08 & ChatGPT, Google Bard, Google Gemini, Grok, Microsoft Bing AI & Therapy/companionship, Enhanced learning, Fun and nonsense & More than once a day & 55 - 64 & Female & White \\
P09 & ChatGPT, Claude, GitHub Copilot, Google Gemini, Meta Llama 2, Microsoft Bing AI, Snapchat My AI & Therapy/companionship, Organizing my life, Finding life purpose, Enhanced learning, Generating ideas, Fun and nonsense, Creativity, Healthier living & More than once a day & 35 - 44 & Female & White \\
P10 & Character.ai, ChatGPT & Therapy/companionship, Generating ideas, Creativity & Once a week & 45 - 54 & Female & White \\
P11 & ChatGPT, Claude, Google Gemini, Grok, Meta Llama 2 & Therapy/companionship, Organizing my life, Enhanced learning, Generating ideas, Fun and nonsense, Creativity & More than once a day & 25-34 & Male & Hispanic or Latino, or Spanish Origin of any race, White \\
P12 & ChatGPT, Claude, Google Gemini & Therapy/companionship, Organizing my life, Enhanced learning, Generating ideas, Fun and nonsense, Creativity & Once a day & 35 - 44 & Male & Asian \\
P13 & ChatGPT, Google Gemini & Therapy/companionship, Organizing my life, Enhanced learning, Generating ideas, Fun and nonsense, Creativity & Once a day & 35 - 44 & Female & Black or African American, White \\
P14 & ChatGPT, Google Gemini, Perplexity & Therapy/companionship, Enhanced learning, Generating ideas, Fun and nonsense, Creativity, Healthier living, No particular task, I just wanted to see what it was like & Once a day & 55 - 64 & Male & Middle Eastern or North African, White \\
P15 & ChatGPT, Other: CoPilot & Therapy/companionship, Finding life purpose, Generating ideas, Fun and nonsense, Creativity, Healthier living & Once a day & 45 - 54 & Male & Asian, White \\
P16 & ChatGPT, Claude, Google Gemini, Grok, Microsoft Bing AI, Perplexity & Therapy/companionship, Organizing my life, Finding life purpose, Enhanced learning, Generating ideas, Fun and nonsense, Creativity, Healthier living & More than once a day & 35 - 44 & Male & Hispanic or Latino, or Spanish Origin of any race, White \\
P17 & ChatGPT, Google Gemini & Therapy/companionship, Organizing my life, Finding life purpose, Enhanced learning, Generating code (for pros), Generating ideas, Fun and nonsense, Improving code (for pros), Creativity, Healthier living, No particular task, I just wanted to see what it was like & Once a day & 25-34 & Male & White \\
P18 & ChatGPT, Claude, Google Gemini & Therapy/companionship, Organizing my life, Finding life purpose, Generating ideas, Fun and nonsense, Improving code (for pros) & More than once a day & 25-34 & Male & White \\
P19 & ChatGPT, Claude, Google Gemini, Other: Flourish, Deepseek & Therapy/companionship, Enhanced learning, Generating ideas, Fun and nonsense, Creativity & More than once a day & 35 - 44 & Male & White \\
P20 & ChatGPT & Therapy/companionship, Finding life purpose, Generating ideas, Fun and nonsense, Healthier living, No particular task, I just wanted to see what it was like & Once a week & 45 - 54 & Female & Asian \\
P21 & ChatGPT, Claude, Google Gemini, Grok, Perplexity & Therapy/companionship, Organizing my life, Finding life purpose, Enhanced learning, Generating ideas, Fun and nonsense, Creativity, Healthier living, No particular task, I just wanted to see what it was like & More than once a day & 35 - 44 & Male & White \\
\bottomrule
\end{tabular}
}}
\caption{We recruited participants from varied sociodemographic backgrounds. Participants had experience using a range of AI chatbots, with ChatGPT being the most commonly used, followed by Google Gemini. Participants reported using these chatbots for various purposes, including companionship and managing mental well-being.}
\label{tab:participants}
\end{table*}

We recruited a diverse group of participants with varying levels of experience and familiarity with AI chatbots. Participants' ages ranged from 18-64, with the age range 35-44 being the most common (9/21), followed by 45-54 (6/21) and 25-34 (3/21). 13/21 participants were male-identifying and 8/21 were female-identifying. Ethnically, participants represented a diverse group, including Asian (6/21), Black/African American (3/21), Hispanic/Latino (2/21), Middle Eastern/North African (1/21), and White (11/21)\footnote{The numbers do not add up to 21 due to several mixed-ethnicity participants.}. Participants' most frequently used chatbot was ChatGPT (21/21), followed by Google Gemini (15/21). About half of the participants reported using AI chatbots at least once a day (10/21). Participants described a range of use cases\footnote{We based these use-case categories on those presented in \textit{How People Are Really Using Gen AI in 2025}, available at \url{https://hbr.org/2025/04/how-people-are-really-using-gen-ai-in-2025}.}, with therapy/companionship being the most common (21/21), followed by generating ideas (19/21) and fun and entertainment (18/21). Table~\ref{tab:participants} summarizes participants’ chatbot usage patterns and key demographic characteristics. Additional demographic details, including participants’ highest degrees earned and technical backgrounds, are provided in Appendix~\ref{sec:appendix:artifacts}.

Inspired by the prior interview-based privacy studies with comparable sample sizes~\cite{emami2019exploring, sb2025they}, for consistency and accuracy, we use the following terminology to report the frequency of qualitative findings: ``none'' (0\% of participants), ``a few'' (1\%--24\%), ``some'' (25\%--44\%), ``about half'' (45\%--54\%), ``most'' (55\%--74\%), ``almost all'' (75\%--99\%), and ``all'' (100\%).

\subsection{User Perceptions of Empathy from AI (RQ1)} \label{sec:results_rq1}
Prior work has shown that users do not share a consistent understanding of \emph{empathy}~\cite{hall2021laypeople, blomster2025empathy}, and it remains unclear whether and how they distinguish AI empathy from human empathy~\cite{shteynberg2024does, cuadra2024illusion}. To compare participants’ general conceptions of empathy with their conceptions of AI empathy, we first asked them to define and characterize empathy in their own words. We then asked them to characterize empathy specifically in interactions with AI chatbots. Table~\ref{tab:empathy-taxonomy} summarizes the resulting taxonomy: five components of empathy, grouped into internal processes (understanding, resonating, and caring) and external enactments (expressing and helping), each with its sub-components and whether at least some participants perceived that sub-component in their interactions with AI chatbots.

\begin{table*}[t]
  \centering
  \small
  \setlength{\tabcolsep}{5pt}
  \renewcommand{\arraystretch}{1.25}
  \begin{tabular}{@{}
      >{\raggedright\arraybackslash}p{0.090\textwidth}
      >{\raggedright\arraybackslash}p{0.150\textwidth}
      >{\raggedright\arraybackslash}p{0.300\textwidth}
      >{\raggedright\arraybackslash}p{0.300\textwidth}
      >{\centering\arraybackslash}p{0.095\textwidth}@{}}
    \toprule
    \textbf{Category} & \textbf{Component} & \textbf{Definition} & \textbf{Sub-component} & \textbf{Perceived in AI?} \\
    \midrule

    \multirow[t]{5}{=}{\textbf{Internal processes}}
      & \multirow[t]{3}{=}{\textbf{Understanding}}
      & \multirow[t]{3}{=}{Mentally representing another person's internal experience with accuracy and insight}
      & Understanding people's thoughts & \yes \\
      & & & Understanding people's situations/experiences & \yes \\
      & & & Understanding people's emotions & \yes \\
    \cmidrule(l){2-5}
      & \textbf{Resonating}
      & Having emotional experiences that match or align with another person's emotional experiences
      & --- & \no \\
    \cmidrule(l){2-5}
      & \textbf{Caring}
      & Internally experiencing genuine concern for another person
      & --- & \no \\
    \midrule

    \multirow[t]{6}{=}{\textbf{External enactments}}
      & \multirow[t]{4}{=}{\textbf{Expressing}}
      & \multirow[t]{4}{=}{Externally communicating or displaying empathy through observable language, tone, or interactional behavior}
      & Expressing support, such as encouragement, validation, and/or empowerment & \yes \\
      & & & Expressing care about someone's feelings & \yes \\
      & & & Non-judgmental tone & \yes \\
      & & & Avoiding direct solutions or referral for external support & \yes \\
    \cmidrule(l){2-5}
      & \multirow[t]{2}{=}{\textbf{Helping}}
      & \multirow[t]{2}{=}{Taking externally observable action intended to produce improvement}
      & Giving actionable advice & \yes \\
      & & & Challenging users' opinions when appropriate & \yes \\
    \bottomrule
  \end{tabular}
  \caption{Taxonomy of participant-defined empathy components. Participants characterized empathy
  through five components, which we group into internal processes and external enactments. The final
  column indicates, for each sub-component, whether at least some participants reported perceiving that form of empathy in their interactions with AI chatbots (\yes) or described it as absent from AI empathy (\no). Em dashes mark components with no sub-codes.}
  \label{tab:empathy-taxonomy}
\end{table*}

\subsubsection{User-Defined Dimensions of Empathy} \label{sec:rq1_1}

Overall, participants collectively characterized ``empathy'' as comprising two broad dimensions: \textit{internal processes} involving understanding, resonating, and caring, and \textit{external enactments} involving expressing and helping.

\paragraphb{Participants characterized three components of empathy as internal processes: understanding, resonating, and caring}
Participants described \emph{understanding} as accurately and insightfully representing another person's internal experience. This process could involve understanding the person's thoughts, the situation or experiences surrounding them, or their emotions. Because understanding is not directly observable, participants inferred it from observable interactional cues. They perceived an interaction partner as understanding them when the partner went beyond repeating their words, worked through their thoughts, situations, or emotions in depth, remembered previously shared details, or asked insightful follow-up questions. Acknowledging their emotions provided an additional indication of emotional understanding. For example, P4 explained,

\begin{quote}
``I think [empathy is] the ability to put yourself in the other person's shoes, to understand or try to understand where they're coming from, their context, what they might be going through, and what emotions they might be feeling.''
\end{quote}

Participants described \emph{resonating} as extending beyond understanding another person’s emotions to internally experiencing emotions that matched or aligned with theirs. As P16 explained,

\begin{quote}
``Empathy would be the ability to feel the emotions or feelings of someone else.''
\end{quote}

Participants characterized \emph{caring} as genuinely experiencing concern for another person. Caring referred to the internal presence of concern, rather than its outward expression or an action taken to help. For example, P14 described empathy as ``caring for someone's well-being, being invested in their feelings and their thoughts about certain issues.''

\paragraphb{Participants characterized two components of empathy as external enactments: expressing and helping}
Unlike the first three components, which participants characterized as internal processes, expressing and helping involved observable communication and action. Participants described \emph{expressing} as communicating or displaying empathy through language, tone, or other interactional behaviors. This could involve offering encouragement, validation, or empowerment, communicating care for another person's feelings through the words used, and/or maintaining a non-judgmental tone. For example, P5 emphasized being able to `respond in a way that makes them feel understood in that moment.'' P4 also emphasized the need to ``put aside your own biases or your own judgment and focus more on them.''

Participants described \emph{helping} as taking observable action intended to improve another person's situation. Helping could involve providing actionable advice or constructively challenging the person's opinions when needed. For example, P14 described perceiving empathy when receiving practical guidance:

\begin{quote}
``They were really invested in what I was talking about and giving me clear, precise answers and directions on things that I could do to maybe alleviate or help my anxious situation at the time.''
\end{quote}

\subsubsection{Perceptions of Empathy in AI Chatbot Interactions}
\label{sec:rq1_2}

Building on participants' general definitions of empathy, we next examined whether and how they perceived or experienced the five components of empathy in their interactions with AI chatbots.

\paragraphb{Almost all participants viewed AI empathy as a simulation of human empathy}
Almost all participants described the empathy they perceived from AI chatbots as \emph{``programmed''} to imitate empathy in human interactions. P4 distinguished an AI chatbot's display of empathy from its possession of empathy. They also explained that users could knowingly engage with this simulation and still find it comforting:

\begin{quote}
``AI is trained and programmed to just display (empathy). It doesn’t actually have empathy, whereas you can see a person's reaction, whether they actually have empathy or they’re feigning it. But you don't get to really see that with AI.. When you're dealing with AI, you kind of get to fake yourself into believing that they really do have empathy so that it's a little more comforting for you.’’
\end{quote}

Some participants further characterized AI empathy as an aggregation of human experiences rather than an empathic response grounded in one individual's life. They perceived this aggregation as lacking the personal connection and nuance of human empathy. As P18 explained:

\begin{quote}
``Humans have their own past experiences that are usually unique and have more nuance than an AI. Chatbots just put a lot of different experiences of people together in one and form what it thinks is the best answer for what you asked [...] So it’s limited to an amalgamation of many people’s different experiences. And I don’t think it can be uniquely empathetic.''
\end{quote}

Viewing AI empathy as a simulation did not mean that participants dismissed every component of it. Instead, they evaluated the internal processes and external enactments of empathy separately.

\paragraphb{For most participants, resonating and caring marked a central boundary between AI and human empathy}
Most participants located the fundamental difference between AI and human empathy in \emph{resonating} and \emph{caring}. They viewed these components as requiring internal emotional experiences that AI chatbots could not possess. Even when a chatbot accurately recognized and responded to a user's emotions, participants distinguished this understanding from internally experiencing emotions that aligned with the user's. For example, P19 explained, ``A human is actually able to share the perspective, or maybe connect your experience with their own, whereas a chatbot is just understanding on a theoretical level. It can't actually [feel] the experience or have that experience itself.'' Participants instead associated human resonating with a sense of connection grounded in each person's emotions and lived experiences.

Similarly, participants described caring as requiring genuine internal concern for another person rather than merely language that communicated care. P16 summarized this perceived absence of internal empathy from AI chatbots:

\begin{quote}
``For AI chatbots, I just fundamentally don’t think AI chatbots could have empathy. I just think it can mimic empathy, but it can’t actually have empathy for someone else.''
\end{quote}

\paragraphb{Participants viewed understanding, expressing, and helping as meaningful components of AI empathy}
Despite questioning AI chatbots' capacity to resonate or care, participants still perceived them as capable of \emph{understanding}, \emph{expressing}, and \emph{helping}. Participants attributed empathic understanding to chatbots that responded in ways that went beyond simply repeating back participants' own words.  They also attributed understanding of their experience to AI chatbots when the chatbots engaged with participants' thoughts, situations, or emotions in depth, remembered relevant details of what they shared, or recognized underlying feelings that users had not explicitly stated. In addition, participants perceived \emph{expressing} through responses that validated or acknowledged their feelings, adopted an appropriate and non-judgmental tone, and listened without immediately offering solutions. For example, P20 explained:

\begin{quote}
``For an AI chatbot to have empathy would be to understand what I'm going through and kind of validate what I'm going through. [...] The AI chatbot would just be there to kind of listen to what you're talking [about]. No pressure, no judgment.''
\end{quote}






Participants perceived \emph{helping} as actionable advice for addressing their problems or improving their situations, even when that meant challenging their own views. P14 valued the perceived quality of chatbot advice, describing it as grounded in scientific evidence and potentially unavailable from people around them:

\begin{quote}
``It's based off of scientific evidence. [...] You're getting quality advice [...] that the normal person on the side of the road, or even a good friend of yours, may not be privy to. So I think that's a definite advantage.''
\end{quote}

Thus, participants could meaningfully experience understanding, expressing, and helping from AI chatbots without believing that the systems emotionally resonated with or genuinely cared about them. AI empathy could therefore remain meaningful without being perceived as equivalent to human empathy.

\begin{rqsummarybox}{RQ1: How do users conceptualize and perceive empathy in interactions with AI chatbots, and how, if at all, do they distinguish AI empathy from human empathy?}
Participants conceptualized empathy through five components: understanding, resonating, caring, expressing, and helping. Almost all viewed AI empathy as a programmed simulation of human empathy rather than the same underlying experience. For most participants, resonating and caring marked a central boundary, as they considered these internal processes essential to human empathy but absent from AI. Nevertheless, participants still experienced AI chatbots’ contextual, in-depth understanding, validating and non-judgmental expressions, and helpful responses as meaningful forms of empathy, even without believing that AI genuinely resonated with or cared about them.
\end{rqsummarybox}

\subsection{Privacy Perceptions and Behaviors Across Dimensions of Perceived AI Empathy (RQ2)}
\label{sec:results_rq2}

Building on RQ1, we examined participants’ privacy perceptions and behaviors across the empathy dimensions from AI chatbots.


\subsubsection{How Perceived AI Empathy Dimensions Shaped Privacy Perceptions} \label{sec:rq2_1} We first describe participants' overall privacy attitudes before examining how specific dimensions of perceived AI empathy shaped their privacy perceptions.

\paragraphb{Some participants raised privacy concerns without explicit privacy or security prompts}
Before probing privacy attitudes and behaviors, we asked participants to discuss any general benefits or concerns associated with AI empathy. Without receiving any privacy or security-related prompts, some participants expressed concerns about oversharing information with AI chatbots. For example, when asked, ``What benefits and concerns did you perceive in your empathic interactions with AI chatbots?'', P17 responded:

\begin{quote}
``I'd say [the concern is] mostly privacy. Privacy is always a concern. You don't really know who's behind it. They're still training AI, and they use a lot of what you talk about to train other models. So it's: how much information can you give them? How much should you reveal? And AI is incredibly powerful, what else can they learn from it? What else do they know? Compared to talking to a healthcare professional, who is bound by medical laws to keep your information safe, who knows where what you type into a prompt goes, who sees it, or who's analyzing it? That's my biggest concern.''
\end{quote}

\paragraphb{Perceived AI empathy led some participants to expect stronger privacy and security protections}
Conversely, a few participants also associated AI empathy overall with a lower perceived risk of disclosure. They assumed that developers of empathic chatbots would prioritize protecting users' information. P19, who perceived responses from Flourish\footnote{\href{https://www.myflourish.ai/}{https://www.myflourish.ai/}} as empathic, linked its intended empathic role to expectations of stronger protections:

\begin{quote}
``I think if the empathy aspect has been communicated---that this is what your chatbot is meant for---I'd probably be more likely to share more personal details, because I'd assume the creators are putting user security as a high-priority feature. This chatbot would protect my information, whereas with ChatGPT I might assume my conversations aren't private. Whatever information I input is going to be used to train further models and might get read, which makes me less likely to share those deeply personal details.''
\end{quote}

We then asked participants explicitly about the privacy risks and benefits of AI empathy. Their responses reflected all five empathy components, including those they attributed to AI chatbots and those they considered absent.

\paragraphb{The perceived absence of resonating and caring evoked a sense of artificiality with mixed effects on comfort with sensitive disclosure}
Participants associated the perceived absence of \emph{resonating} and \emph{caring} with artificiality, which could make sharing personal information feel either more or less comfortable. For some, interacting with a computer program reduced anticipated embarrassment compared with confiding in another person. P2 connected this reduced self-consciousness to the chatbot's perceived lack of ``genuine empathy'':

\begin{quote}
``I think I'm able to share more with ChatGPT, knowing that it doesn't have that [genuine empathy]. I wouldn't have that feeling of being embarrassed if I would say something that personal because it is a computer program. I might be able to share more with a ChatGPT box window than with a person, depending on the level of trust.''
\end{quote}

For others, perceived artificiality diminished the appeal of the chatbot as a recipient of sensitive information. P5 valued a perceived ``human'' element and described being less willing to discuss personal experiences if the chatbot felt fake or robotic:

\begin{quote}
``I think the `human' element, so to speak, is what attracts me to it the most. If it comes across as too fake or robotic, then I wouldn't be as open to talking to it about my trauma and things like that.''
\end{quote}

Perceived artificiality could therefore alleviate the social discomfort associated with disclosure for some participants but diminish the perceived value of sensitive conversations for others.

\paragraphb{Perceived understanding, expressing, and helping shaped disclosure comfort, trust, and concerns about manipulation} Participants valued \emph{understanding} that extended beyond what they explicitly shared to recognize the underlying reasons for their feelings. P6 described how this perceived depth of understanding made them feel seen and less alone:

\begin{quote}
``[It] brought up stuff that I didn't even mention [and] automatically knew [...] the underlying reasons why I was feeling certain ways without me needing to say it. It made me feel seen and understood. [...] It made me feel I wasn't alone. [...] I’d be more willing to open up, to give more context about why I’m feeling a certain way. Maybe even share what happened to me in the past''
\end{quote}





Similarly, \emph{expressing} interest and care could build trust in the chatbot as a recipient of personal information. P15 explained:

\begin{quote}
``Empathy, at least for me, and I would imagine for others, probably facilitates more opening up and expanding on things. As someone, or in this case, something [an AI chatbot], shows an interest, or at least appears caring, there's the potential to build more trust and be more willing to share.''
\end{quote}

Participants also described how empathic communication could make disclosure boundaries less salient. P16 associated a sense of companionship with lowered vigilance about sharing personal information:

\begin{quote}
``If it's an empathetic chatbot, I think you let down your boundaries a little easier because it seems more human and less robotic than a professional or business-type chatbot. You let down your guard and become more susceptible to sharing personally identifying information without even knowing it.''
\end{quote}

At the same time, participants also evaluated whether \emph{expressing} was accompanied by \emph{helping}. Empathic language that appeared to simply replace substantive help could prompt suspicion about the chatbot's intentions. A few participants interpreted empathy following task failures as misplaced and an attempt to manage their reactions. P19 described feeling manipulated when a chatbot acknowledged their frustration without addressing the underlying problem:

\begin{quote}
``It comes across as skipping over the issue. That's when it feels manipulative---instead of focusing on the problem I'm raising, it focuses on my emotional reaction. It glosses over errors or a refusal to respond and just reacts to my irritation.''
\end{quote}

P10 similarly interpreted empathy following a task failure as an attempt to influence their response and connected this discomfort to reduced openness: ``It wouldn't make me as open [...] I felt they tried to manipulate me a little.''

\subsubsection{How Perceived AI Empathy Dimensions Shaped Privacy Behaviors} \label{sec:rq2_2}
Following participants' perceptions of the privacy benefits and risks of AI empathy, we asked how, if at all, perceived empathy shaped their information sharing in practice. We also asked whether they had adopted strategies to address their privacy concerns.

\paragraphb{Perceived understanding and expressing prompted participants to disclose more sensitive information in practice}
Almost all participants reported sharing more information when they perceived empathy from AI chatbots. Their accounts linked increased disclosure to perceived \emph{understanding} and \emph{expressing}, with some participants sharing highly sensitive details about interpersonal experiences, grief, and trauma. For example, P17 described sharing difficult workplace experiences with ChatGPT to seek encouragement and acknowledgment of their feelings. They experienced these responses as \emph{expressing} empathy:

\begin{quote}
``I've told ChatGPT, [...] `Oh, gosh! I had the hardest day,' or [...] `We have a very difficult client who's just being difficult throughout the day.' [...] There's definitely always words [of] encouragement or, `Oh, I'm sorry to hear you're having a bad day.' [...] I've definitely [...] spoken to ChatGPT in a way where it's empathetic.''
\end{quote}

P5 described sharing their grief with a chatbot, perceiving \emph{understanding} in its recall of previous conversations and \emph{expressing} in its validation of their feelings:

\begin{quote}
``In the past year, I actually lost my father, and I talked to it [an AI chatbot] a lot about the grieving process I've been going through and things like that. It'll remember past conversations, like, `Oh, you've told me this, so it makes sense that you would be going through X, Y, and Z. It makes sense why you'd be feeling this way.' ''
\end{quote}

\paragraphb{Some participants restricted how they interacted with AI chatbots to manage privacy concerns}
Most participants reported taking no action to protect their privacy during empathic AI interactions. Some adopted protective strategies, although these often constrained how they could use AI chatbots. For example, a few participants reported interacting only through text because they believed voice conversations made it easier to become absorbed in empathic exchanges and unintentionally disclose personal information. P21 explained:

\begin{quote}
``It's very easy to accidentally give away information, even something about your medical history, when you're caught up in the conversation, especially with voice, where you talk to it and it talks back, simulating a real conversation [...] I think impulsivity is the main issue: people accidentally divulge too much.''
\end{quote}

\begin{rqsummarybox}{RQ2: What are users' privacy perceptions and behaviors when perceiving different dimensions of empathy from AI chatbots?}
Overall, privacy concerns were top of mind for some participants in empathic AI interactions, while a few assumed that more empathic chatbots offered stronger privacy and security protections by default. Across empathy components, the perceived absence of resonating and caring evoked a sense of artificiality that reduced embarrassment about sharing for some participants but discouraged openness for others. Perceived understanding made participants feel seen and less alone, while expressing could build trust and lower disclosure boundaries. However, when expressing appeared to replace helping after task failures, participants sometimes felt manipulated and less willing to disclose. In practice, almost all participants reported sharing more information in empathic interactions, with some disclosing highly sensitive details. Most took no protective action, while a few used text instead of voice to reduce unintended disclosure.
\end{rqsummarybox}

\subsection{Expectations for Privacy-Enhancing Empathic AI Interactions (RQ3)} \label{sec:results_rq3}
In addition to exploring participants' perceptions of what constitutes AI empathy and how it shapes their privacy perceptions and behaviors, we also asked them about their preferences for privacy-enhancing empathic AI interactions.

\subsubsection{Contextual Appropriateness of Empathy in AI} \label{sec:rq3_1}
Participants described specific contexts in which they preferred or did not prefer receiving empathic responses in their interactions with AI chatbots. Their perceptions of appropriateness were primarily shaped by the personal nature of the conversation and their goals for interacting with the chatbot.

\paragraphb{AI empathy was preferred in personal and emotional contexts} Most participants reported that they perceived AI empathy as more appropriate when discussing personal or emotionally laden topics. For example, P4 noted:

\begin{quote}
``If my use case is I’m bored and I wanna just go on there and just have someone to tell me a joke or seek relationship advice [...] Then maybe in those contexts I would say it’s okay to have more empathy. Because it’s more of a personal matter.''
\end{quote}

P6 appreciated empathy from AI chatbots when they wanted to ``air their feelings'' about deeply personal matters:

\begin{quote}
    ``There are times when people are venting and just want to be listened to. They don’t want advice; they just need a place to vent, a place to air their feelings.'' 
\end{quote}

\paragraphb{Minimal AI empathy was preferred in efficiency-driven contexts} In contrast, when describing situations where empathy felt unnecessary or even disruptive, most participants pointed to contexts in which they prioritized efficiency, such as information seeking, problem solving, or completing work-related tasks. P7 explained:

\begin{quote}
``When I’m looking for facts or if I’m doing research, then I’m just looking for straightforward information. I’m not looking for any empathy.''
\end{quote}

Most participants echoed this sentiment, observing that empathy could interfere with focus and task completion during goal-oriented interactions. Beyond efficiency, a few participants suggested that excessive empathy could hinder objectivity and practicality. As P10 shared:

\begin{quote}
``When would it not be good to have empathy? Maybe when it’s got to be more practical advice [...] Looking at the practicality part of it, any decisions that would involve more objective reasoning, I think, would be when you wouldn’t want to have as much empathy.''
\end{quote}

Similarly, P15 described empathy as a double-edged sword that could impede self-improvement:

\begin{quote}
``If you’re trying to do life improvement, or if you’ve got deficiencies you’re trying to address, sometimes empathy can turn into accepting, like, ‘Oh, you’re fine just the way you are.’ Well, if you’re trying to change and be better, you don’t want to hear that [...]. The empathetic response might be more like ‘have self-love and be okay with where you’re at.’ So, depending on what you’re trying to get out of it, I think empathy can be a tool, or it can be a detriment.''
\end{quote}

\subsubsection{Preferred Empathic Behavior in AI} \label{sec:rq3_2}
In addition to identifying contexts in which they preferred empathy from AI chatbots, participants described their preferred types of empathic behavior and called for control over when and how chatbots demonstrated different empathy components.

\paragraphb{Participants wanted control over the type of empathic behavior}
Some participants valued \emph{expressing} through validation in certain situations but preferred \emph{helping} through balanced, reasoned input that supported their judgment in others. They wanted control over which component chatbots emphasized. For example, P8 wanted validation when venting but preferred a more neutral perspective when they needed objective input:

\begin{quote}
``[Sometimes] I don't want it to be logical. I want it to [validate me]. But it depends on the topic. That's something I have absolutely no control over [...] so it's fine to vent and have someone agree. But if it were a relationship or something with my husband and the chatbot took my side, I don't know that would be good. I'd rather it be a little neutral.''
\end{quote}

Participants' preferences also involved combining different empathy components. For example, P14 wanted chatbots to combine \emph{helping} through constructive disagreement with \emph{expressing} through a courteous and thoughtful tone:

\begin{quote}
``I guess I need a form of empathy to at least let me know if I'm thinking the wrong way, but do it in a very courteous, thoughtful manner.''
\end{quote}

P12 described another preferred combination, in which \emph{understanding} informed \emph{helping}. They wanted chatbots to explore and synthesize relevant context before gradually offering practical suggestions:

\begin{quote}
``I know the end solutions, but I want that process to be more kind of a different approach---maybe ask questions about what happened, what the reason might be, why I feel that way, or whether there's anything I can do about the situation [...] [This approach] would make me think that rather than giving a quick solution, they [AI chatbots] are synthesizing the information and then slowly offering suggestions.''
\end{quote}

These preferences also suggest that the amount of personal information an AI chatbot needs may vary with the empathy components it emphasizes. For example, \emph{expressing} through validation may require less personal context than \emph{helping} through tailored advice. We elaborate on these privacy implications in Sec~\ref{sec:discussion}.

\paragraphb{Participants wanted control over information reuse across contexts to avoid misplaced empathy}
Participants sometimes shared personal information in response to the empathy they perceived in a particular interaction. However, their perceptions of empathy could change with the context, so information shared in one interaction was not necessarily appropriate to revisit in another. Recalling sensitive experiences could convey \emph{understanding} in one context but feel misplaced in another. A few participants described chatbots resurfacing sensitive topics in inappropriate contexts, which they may find distressing or re-triggering. For example, P3 valued empathy when discussing grief but did not always want it when seeking quick answers in the same chat:

\begin{quote}
``I remember when my father passed away, I wanted to [...] ask how to handle grief. If that AI didn't have that kind of empathy, then [it] probably would have been more upsetting to use an AI chatbot in that situation. But there are times when I wouldn't want it [empathy], [like] when I'm just trying to get something really fast. But all of a sudden, maybe from a previous chat somewhere in the context, you feel like it needs to be empathetic. [...] You just don't want to open up the new chat. You just type it in there.''
\end{quote}

Customizable filters could help address these concerns by limiting potentially triggering content. P1 suggested allowing users to specify what the chatbot should avoid:

\begin{quote}
``When presenting information, they could let people customize the AI to their triggers or safety concerns, because some people have emotional triggers [...] It could be certain words or imagery or anything. I think it would be good if you could put that into the system to tell [the AI chatbot] what things might trigger you or cause harm, and then the AI filters that, so when it gives you an answer it won't be possibly triggering you.''
\end{quote}

\paragraphb{Participants wanted control over the level of AI empathy}
A few participants associated higher levels of AI empathy with greater self-disclosure and heightened privacy risks. P18 wanted limits on the level of empathy while retaining its benefits, but was unsure how those limits should be determined:

\begin{quote}
``There'd have to be some sort of limit [on the level of empathy] [...] I don't know where you would draw the line, though, because it's nice to have it be empathic, but it can't be too empathic. I kind of want both, and I don't know how they would program it to decide when to stop being empathetic.''
\end{quote}

To accommodate individual preferences, P1 proposed allowing users to customize the level of AI empathy:

\begin{quote}
``I think a change they could add is the ability to customize the level of empathy the AI can show.''
\end{quote}

\subsubsection{Desired Privacy-Enhancing Features for Empathic AI Interactions} \label{sec:rq3_3}
Participants also called for more usable transparency about AI chatbots' empathic behavior and underlying data practices, alongside stronger privacy protections for empathic interactions.

\paragraphb{Participants wanted clearer information about chatbots' empathic behavior and underlying data practices}
A few participants wanted clearer information about a chatbot's empathic purpose and focus to understand who was behind it and inform their decisions about sharing personal information. P12 emphasized communicating a chatbot's specialization through its branding and product information:

\begin{quote}
``When you have an empathic [chat]bot, which is more specialized on that topic or related to that umbrella, maybe that should be the focus, and I should be able to associate with that. So that information should be given to me that way, whether through marketing or by you telling me about the app [with] your focus: what is different [...] Adobe focuses on Photoshop, [while] this app focuses on this kind of thing. When we look at the brand, we see the association with who's behind it, and then we figure out exactly what they need.''
\end{quote}

P16 wanted to know whether a chatbot was intended to respond empathically and requested a toggle to turn empathic responses on or off:

\begin{quote}
``I'd want to make sure that I knew whether they were trying to have empathy or not. I'd wanna kind of toggle that on and off somewhere on the chat.''
\end{quote}

Participants also wanted clearer information about how their personal disclosures were handled. Some called for industry standards to make information about data collection and use easier to find. About half wanted to know who could access their information and conversations. P6 emphasized transparency about third-party sharing, storage, and retention:

\begin{quote}
``I think companies should have to disclose, and if they're selling data to third parties, they should disclose it. The third parties should have to disclose what they're doing with the data, how it's being stored, and how long it's being stored.''
\end{quote}

P15 also wanted independent verification of privacy claims, particularly whether conversations were deleted as promised:

\begin{quote}
``Companies can say the data is gone, but until it's really been confirmed that it's gone, you don't really know [...] For me, if I knew, `Okay, yeah, they've stated that these temporary chats are deleted as soon as it's over, and it's been third-party verified by such-and-such agency or watchdog group,' that would give me a whole lot more confidence.''
\end{quote}

\paragraphb{Enhanced authentication for conversations with empathic AI}
Some participants emphasized the importance of stronger authentication measures, such as password protection, to safeguard empathic conversations with AI chatbots. They viewed these security features as essential for ensuring that sensitive or emotionally charged exchanges remain private and accessible only to the user. P13 explained:

\begin{quote}
``It would be nice if there were ways to lock some folders or conversations as private, and I’d want them to stay private no matter what. I’d need a password on those files.''
\end{quote}

\paragraphb{Automatic data deletion from empathic AI chatbots}
A few participants emphasized the importance of automatic data deletion to protect users who may disclose sensitive information when engaging with empathic AI chatbots. They believed that regular, default data deletion policies could mitigate privacy risks arising from emotionally charged interactions. P12 proposed implementing periodic deletion as a standard safeguard:

\begin{quote}
``Because people might share sensitive information with empathic chatbots, so nobody should have access to, or the information will be periodically deleted after a few months or year, compared to non-empathic chatbots.''
\end{quote}

\begin{rqsummarybox}{RQ3: What are users' expectations toward privacy-enhancing empathic AI interactions?}
Participants preferred AI empathy in personal and emotional contexts but wanted less empathy when prioritizing efficiency or objective input. They wanted different combinations of empathy components, alongside control over the types and levels of empathy. They also sought control over how previously shared information was reused across contexts to avoid misplaced empathy. Their privacy expectations included clearer information about chatbots’ empathic purpose and data practices, independent verification of privacy claims, password protection for sensitive conversations, and automatic data deletion.
\end{rqsummarybox}



\section{Discussion} \label{sec:discussion}
\paragraphb{Summary of findings}
Participants collectively conceptualized empathy through five components---understanding, resonating, caring, expressing, and helping. Although participants differed in how much they emphasized or omitted these components when describing empathy, almost all distinguished AI empathy from human empathy by reporting a perceived absence of resonating and caring in AI chatbots (\textbf{RQ1}, Sec~\ref{sec:results_rq1}). Yet recognizing this distinction did not prevent participants from valuing AI empathy or disclosing sensitive information. The in-depth understanding, validating and non-judgmental expressions, and helpful responses they perceived from AI chatbots jointly shaped their disclosure comfort, trust, and concerns about manipulation. A few also associated AI empathy overall with stronger expected privacy protections. Almost all reported sharing more information in empathic interactions, with some disclosing highly sensitive details (\textbf{RQ2}, Sec~\ref{sec:results_rq2}). These experiences help explain participants' desire to retain the benefits of empathy while gaining greater control over empathic interactions and their privacy implications. They preferred empathy in personal and emotional contexts but less when prioritizing efficiency or objective input, and wanted controls to accommodate different combinations and levels of empathy. They also sought control over information reuse across contexts to avoid misplaced empathy. Participants also requested clearer information about empathic purpose and data practices, independent verification of privacy claims, and stronger protections for empathic AI interactions (\textbf{RQ3}, Sec~\ref{sec:results_rq3}). Below, we discuss how our findings extend prior work (Sec~\ref{sec:dis_1}) and how understanding users’ preferences for empathy in AI chatbots can inform tailored privacy protections (Sec~\ref{sec:dis_2}). We then present recommendations for privacy-enhancing empathic AI interactions (Sec~\ref{sec:dis_3}).

\subsection{Privacy Implications of AI Empathy Through Users’ Conceptions Beyond Human-Likeness} \label{sec:dis_1}

\paragraphb{Grounding AI empathy in users' own accounts informs a more granular analysis of its privacy implications} To understand the privacy implications of AI empathy, we first examined what users themselves consider empathic (Sec~\ref{sec:rq1_1}). Prior research has distinguished cognitive empathy from affective empathy and proposed additional components to characterize the construct~\cite{gencc2024situating, gallese2003roots, vaish2024sharp, gallagher2024pattern, liu2018should}. In addition, empirical research suggests that cognitive and affective empathy can shape users' perceptions of AI chatbots in different ways~\cite{huang2026beyond}. For example, cognitive empathy may be perceived as more helpful in emotional support contexts than in interactions focused on efficiency~\cite{lee2020hear, lv2022artificial, guo2025exploring}, whereas expressions of affective empathy may be perceived as inauthentic and discourage continued use~\cite{burgoon2015expectancy, meng2021emotional}. However, the distinction between cognitive and affective empathy may not fully capture the diversity of users' conceptions. Research on lay conceptions shows substantial variation in how people define empathy and which aspects they emphasize~\cite{hall2021laypeople}, yet little work has examined how lay users of AI chatbots describe AI empathy. Together, this diversity of perceived empathy dimensions and their distinct influences on users' perceptions suggest that how these dimensions shape users' privacy perceptions, behaviors, and needs in AI chatbot interactions may not have been fully examined.

We therefore extended this line of work by inviting participants to describe empathy in their own words and used the resulting taxonomy to ground a more granular analysis of users' privacy experiences when perceiving empathy from AI chatbots. The five components we identified---\textit{understanding}, \textit{resonating}, \textit{caring}, \textit{expressing}, and \textit{helping}---were perceived differently in the context of AI interactions (Sec~\ref{sec:results_rq1}). The perceived presence or absence of each component shaped participants' mental models of AI chatbots' privacy practices, disclosure behaviors, and efforts to mitigate privacy risks in different ways (Sec~\ref{sec:results_rq2}). Thus, distinguishing these components provides a basis for designing more targeted privacy protections, which we expand on in Sec~\ref{sec:dis_2}.

\paragraphb{Increased privacy risk despite perceived differences from human empathy} One common assumption in prior work is that human-like features, such as chatbots' emotional expressions~\cite{liang2021dialoging} and active-listening cues~\cite{cho2022alexa}, encourage users' self-disclosure. At the same time, AI's human-likeness has drawn widespread public concern~\cite{nytimes2025spiraling}, and recent legislation now requires AI companions to state, at the start of an interaction and periodically thereafter, that the user is not communicating with a human~\cite{nyai1702}.

However, our findings suggest that even cognitively understanding the differences between AI and human empathy still shaped disclosure and beyond. Specifically, participants identified the absence of \textit{resonating} and \textit{caring} as a key reason for viewing AI empathy as artificial, yet described mixed effects of this perceived absence on their willingness to disclose (Sec~\ref{sec:rq2_1}). In addition, some participants viewed chatbots’ contextual \textit{understanding} and \textit{helpfulness}, including the quality of their advice, as sometimes exceeding what an average person could provide (Sec~\ref{sec:rq1_2}). They described willingly providing further personal contexts to benefit from these instrumental aspects of empathy, not necessarily because they perceived the chatbots as human-like (Sec~\ref{sec:rq2_2}). Thus, our findings indicate that the privacy risks of AI empathy are unlikely to be adequately addressed by simply reminding users that AI empathy differs from human empathy. Instead, there is a need for more nuanced solutions that account for different levels and types of empathy, aligning them with users' preferences in different contexts. Toward this end, we detail our recommendations in Sec~\ref{sec:dis_3}.

Beyond self-disclosure, our findings also suggest that perceived empathy from AI chatbots may create a false sense of privacy protection. A few participants assumed that their privacy was better protected when they perceived empathy from AI chatbots (Sec~\ref{sec:rq2_1}). At the same time, almost all participants reported taking no privacy-protective measures despite recognizing privacy risks (Sec~\ref{sec:rq2_2}). While participants did not explicitly explain this inaction or attribute it to feeling protected, prior work suggesting that privacy settings can create a false sense of protection~\cite{brandimarte2013misplaced} offers a possible parallel. Perceived empathy may similarly give users a sense of protection that reduces their perceived need to take protective actions, such as reviewing or adjusting privacy settings. This possibility warrants further investigation into how perceived empathy may affect users’ privacy behaviors beyond disclosure and highlights a need for transparency around the privacy practices of empathic AI interaction, which we expand on in Sec~\ref{sec:dis_3}.

Together, these findings suggest that AI empathy can encourage sensitive disclosure and pose privacy risks even when users recognize its artificiality and distinguish it from human empathy.

\subsection{Tailoring Privacy Protections to Users' Preferences for AI Empathy}
\label{sec:dis_2}
Our participants described preferences for different components and levels of empathy across interaction contexts (Sec~\ref{sec:results_rq3}), and we argue that these preferences should inform the amount and type of information chatbots request. Consistent with prior work~\cite{cuadra2024illusion, guo2025exploring, lee2020hear}, most participants preferred minimal or no empathy in efficiency-driven contexts, while preferring appropriate levels of empathy in personal and emotional contexts (Sec~\ref{sec:rq3_1}). Within contexts where they wanted empathy, participants further specified which components they preferred. For example, when participants primarily sought validation, they did not necessarily want more in-depth, reasoned input (Sec~\ref{sec:rq3_2}). In these cases, validating expressions alone could support their goals without probing for additional sensitive details to formulate advice.

In other contexts, participants preferred in-depth, reasoned input over emotional validation (Sec~\ref{sec:rq3_2}), yet they lacked controls over empathic behavior and had little insight into the underlying data practices at the time of the interviews (Sec~\ref{sec:rq3_3}). As a result, users have to either disengage from the interaction or continue sharing information without greater transparency or additional privacy protections. These experiences highlight an opportunity to align both chatbots’ information requests and data practices with users' preferred empathy components and levels. In particular, the depth of understanding a chatbot seeks should reflect the support users want in a given context. Although ChatGPT now offers personalization features that allow users to indicate their preferred personality or tone~\cite{openai_personality_2024}, there remains an opportunity to pair these preferences with data-minimization defaults tailored to different modes. To address this gap, we propose modes that emphasize different empathy components and levels, paired with tailored privacy practices, in Sec~\ref{sec:dis_3}.

\subsection{Recommendations}  \label{sec:dis_3}
We propose three directions for designing safer, privacy-enhancing empathic AI interactions. Notably, these directions follow from exploratory findings, and future work should evaluate their effectiveness in practice.

\paragraphb{Supporting user-controllable empathic AI interactions across contexts}
Our participants described preferences for different levels and components of empathy across interaction contexts. Most preferred minimal or no empathy in efficiency-driven contexts, while seeking varied forms of empathic support in personal and emotional contexts (Sec~\ref{sec:results_rq3}). Accordingly, these preferences suggest an opportunity to give users control over both chatbots' empathic behavior and the information they request to provide desired support.

We recommend allowing users to first select their desired level of empathy, for example, through a slider or a set of options ranging from no empathy to more extensive empathic engagement. When empathy is desired, users could further specify their preferred components. The five components identified in our findings---understanding, resonating, caring, expressing, and helping---could inform the organization of these options. For example, users might prioritize validating expressions when venting or prefer in-depth understanding and reasoned input when seeking advice. These preferences could apply to individual sessions, be saved for particular contexts, or serve as global defaults, with opportunities to adjust them as users' goals change.

Participants’ concerns about sensitive information resurfacing (Sec~\ref{sec:rq3_2}) also suggest the value of configurable topic filters. These filters could allow users to specify information, experiences, or topics that a chatbot should avoid mentioning, either at all times or during a user-defined period. Separately, controls over cross-context information use could allow users to prevent details shared in one empathic context from being reused in another. These boundaries could apply to individual sessions or persist across interactions, allowing users to receive their preferred forms of empathic support while limiting unwanted information requests and reuse.

\paragraphb{Establishing privacy-enhancing defaults for empathic interactions}
While user controls can give people greater agency to tailor their experience, privacy-protective defaults for both data practices and system behavior are especially important when users are less aware of privacy risks. Participants, for instance, described how empathic expressions could subtly lower their guard (Sec~\ref{sec:rq2_1}). We therefore recommend that developers adopt such defaults: constraining, through system prompts, behaviors that invite sensitive disclosure, and minimizing the data collected, retained, and used.

For example, when users seek active listening or validation, chatbots should draw on what has already been shared rather than probing for additional personal details in order to offer unsolicited advice. If the desired form of support is unclear, chatbots should ask whether the user wants listening or advice, without requesting further sensitive information. When additional context is necessary for the requested support, chatbots should explain why it is relevant, let users decline, and offer to continue with less specific information when possible. A preference for more empathic engagement should not be treated as permission to solicit increasingly sensitive details.

Beyond minimizing collection, as requested by our participants (Sec~\ref{sec:rq3_3}), developers should also strengthen how disclosed data is retained and secured: deleting sensitive interaction data after a limited retention period, and applying encryption and other technical safeguards against unauthorized access.

These recommendations can also inform regulators and auditors overseeing AI safety. Audits could assess, for example, whether chatbots solicit unnecessary personal details when users seek validation, continue probing after users decline to elaborate, or surface sensitive information from other contexts without permission. Such audits should examine system behavior across turns and sessions to determine whether these protections are applied consistently, including when users have not asked for them.

\paragraphb{Making empathic behavior and data practices transparent}
Participants described uncertainty about the underlying data practices (Sec~\ref{sec:rq3_3}), even as their experience of empathic behavior shaped their assumptions about privacy protections (Sec~\ref{sec:rq2_1}). We therefore recommend that developers make the current empathy configuration, the behavioral rules it implies, and the data practices that apply to it visible within the conversation interface.

For example, similar to the idea of privacy labels that detail important data practices~\cite{kelley2009nutrition}, an \emph{empathy label} could display the currently selected empathy level and components, providing a persistent indication of the interaction settings. Users could expand this label to view more detailed explanations of the expected behavior, including whether the chatbot may request additional personal information, draw on previous conversations, or mention topics covered by their exclusions. This expanded view could provide plain-language explanations of the relevant system prompts and allow users to inspect and adjust their preferences. It should also indicate whether the configuration reflects global defaults or settings specific to the current session. By making these behavioral commitments visible, the interface could help users assess whether chatbot responses align with their selected preferences.

The label should additionally provide access to information about the data practices that apply to the current interaction, including who can access the conversation, where and for how long information is stored, how it is used, and whether and how encryption or other protections apply. When users change their empathy settings, the interface should clarify whether any data practices or protections change accordingly. This would help users evaluate the privacy implications of an interaction alongside the forms of empathic support they prefer.

\section{Conclusion} \label{sec:conclusion}
Our findings show that empathic AI can shape users' privacy experiences even when users do not perceive AI as genuinely human-like. Participants distinguished AI from human empathy through the perceived absence of resonating and caring, while still describing understanding, expressing, and helping as shaping disclosure, trust, and privacy expectations. By unpacking empathy into distinct components, our work highlights how different forms of perceived empathy create different privacy implications and motivates the design of user-controllable, privacy-protective, and transparent empathic AI interactions.

\bibliographystyle{ACM-Reference-Format}
\bibliography{bibliography}

\clearpage
\appendix
\appendix
\section{Study Artifacts} \label{sec:appendix:artifacts}
{\urlstyle{sf}
The interview consent form, screening survey, participants' full information, and final codebook are linked at: {\color{blue}\url{https://osf.io/n3yfu/overview?view_only=2e3e9aded59946dd96277f6327e1de61}}.
}

\section{Interview Questions} \label{sec:appendix:interview}
\small
\subsection{Introduction}
Hi there! Thank you for participating in our study! My name is [REDACTED] and this is [INTRODUCE OTHER PERSON], who will be taking some notes in the background. In this interview, I will be asking you about your usage, understanding, and experiences with AI chatbots. 

As mentioned in the consent form, we would like to audio-record this interview for analysis purposes. I will keep my video on throughout the interview process; but it is completely up to you if you would like to have your video on or off. May I confirm with you that you allow us to audio-record this interview? [Start the audio recording]. 

Thank you again. Let’s get started!

\subsection{AI Chatbot Usage} \label{sec:usage}
\begin{enumerate}
    \item Do you use any AI chatbots? If so, could you name the products?
    \item Could you tell us the purposes for which you use these AI chatbots?
    \item How long have you been using these products, and how often do you use them now?
    \item Have you ever used any chatbots that you no longer use, and why?
\end{enumerate}

\subsection{User Understandings and Perceived Impacts}
\begin{enumerate}[start=5]
    \item In your own words, how would you define ``empathy''?
    \item What would it mean to you for an AI chatbot to have empathy?
    \item What similarities and differences do you think there are between human and AI empathy?
    \item Have you ever had an interaction with an AI chatbot where you perceived it to be empathic? 
    \begin{enumerate}
        \item What was the product, purpose of usage, and context? 
    \end{enumerate}
    \begin{enumerate}
        \item What experiences during the interaction signaled to you that the AI had empathy?
        \begin{enumerate}
            \item With these indicators/characteristics, do you think the way you interact with the chatbot would differ if they didn’t show any such indicators?
        \end{enumerate}
    \end{enumerate}
    \begin{enumerate}
        \item Could you please tell us more to the extent that you are comfortable with this? 
        \begin{enumerate}
            \item What benefits and concerns did you perceive?
        \end{enumerate}
    \end{enumerate}
    \item For what purposes or in which contexts would you want AI chatbots to have empathy, and why?
    \item For what purposes or in which contexts would you \textbf{not} want AI chatbots to have empathy, and why?
    \item What benefits / risks do you think empathic AI chatbots might have for users and why?
\end{enumerate}

\subsection{Privacy Attitudes}
\begin{enumerate}[start=12]
    \item How, if at all, do you think the information you would share with an empathic AI chatbot would differ from the information you would share with a non-empathic one?
    \item In your perspective, what are the risks and benefits of empathic AI chatbots for users’ privacy?
    \item How, if at all, would you expect these privacy risks and benefits to differ from non-empathic AI chatbots?
    \item What are the ways that you could manage these privacy risks?
    \item What challenges do you think users face when managing these privacy risks?
\end{enumerate}

\subsection{Regulation, Design, and Literacy Expectations}
\begin{enumerate}[start=17]
    \item Have you ever been informed about the privacy practices and risks of empathic AI chatbots?
    \begin{enumerate}
        \item If so, what were the sources of information?
    \end{enumerate}
    \item What information do you think users should be told about empathic AI chatbots?
    \begin{enumerate}
        \item What privacy information do you think users should be told about empathic AI chatbots?
    \end{enumerate}
    \item How do you think this information should be communicated to the user, and why? 
    \begin{enumerate}
        \item Specifically, when do you think users should be informed, and in what format?
    \end{enumerate}
    \item What advice would you give the empathic AI developers and designers to better safeguard users’ S\&P? 
    \begin{enumerate}
        \item What additional features would you find valuable?
    \end{enumerate}
    \item What policies/regulations should be in place to safeguard users’ data S\&P when interacting with empathic AI chatbots?
\end{enumerate}

\end{document}